\documentclass[12pt]{amsart}
\usepackage{amssymb}
\newtheorem{thm}{Theorem}
\newtheorem{prop}{Proposition}

\newtheorem{cor}{Corollary}
\newtheorem{defn}{Definition}
\newcommand{\intprod}{\makebox[10pt]{\rule{5pt}{.3pt}\rule{.3pt}{5pt}}}

\newcommand{\im}{\operatorname{\mathrm{im}}}
\newcommand{\topten}{\circledcirc}
\begin{document}

\title{Higher Symmetries of the Laplacian}
\author[Michael Eastwood]{Michael Eastwood}
\address{Department of Pure Mathematics\\ University of Adelaide,\newline
         \indent South AUSTRALIA 5005}
\email{meastwoo@maths.adelaide.edu.au}

\begin{abstract} We identify the symmetry algebra of the Laplacian on
Euclidean space as an explicit quotient of the universal enveloping algebra of
the Lie algebra of conformal motions. We construct analogues of these
symmetries on a general conformal manifold.
\end{abstract}

\maketitle

\renewcommand{\thefootnote}{}
\footnotetext{Support from the Australian Research Council is gratefully
acknowledged.}

\section{Introduction}
The space of smooth first order linear differential operators on
${\mathbb R}^n$ that preserve harmonic functions is closed under Lie bracket.
For $n\geq 3$, it is finite-dimensional (of dimension $(n^2+3n+4)/2$). Its
commutator subalgebra is isomorphic to ${\mathfrak{so}}(n+1,1)$, the Lie
algebra of conformal motions of~${\mathbb R}^n$. Second order symmetries
of the Laplacian on ${\mathbb R}^3$ were classified by Boyer, Kalnins, and
Miller~\cite{bkm}. Commuting pairs of second order symmetries, as observed by
Winternitz and Fri\v{s}~\cite{wf}, correspond to separation of variables for
the Laplacian. This leads to classical co\"ordinate systems and special
functions~\cite{bkm,miller}.

General symmetries of the Laplacian on ${\mathbb R}^n$ give rise to an algebra,
filtered by degree (see Definition~\ref{definitionofAn} below). For $n\geq 3$,
the filtering subspaces are finite-dimensional and closely related to the space
of conformal Killing tensors as in Theorems~\ref{Killing} and~\ref{existence}
below. The main result of this article is an explicit algebraic description of
this symmetry algebra (namely Theorem~\ref{structure} and its
Corollary~\ref{Aasalgebra}). Most of this article is concerned with the
Laplacian on~${\mathbb R}^n$. Its symmetries, however, admit conformally
invariant analogues on a general Riemannian manifold. They are constructed in
\S\ref{curved} and further discussed in~\S\ref{discuss}.

The motivation for this article comes from physics, especially the recent
theory of higher spin fields and their symmetries: see \cite{mik,ss,v} and
references therein. In particular, conformal Killing tensors arise explicitly
in \cite{mik} and implicitly in \cite{v} for similar reasons. Underlying this
progress is the AdS/CFT correspondence~\cite{gr,ma,wi}. Indeed, we shall use a
version of this correspondence to prove Theorem~\ref{existence} in
\S\ref{flatresults} and to establish the algebraic structure of the symmetry
algebra in \S\ref{algebraicstructure}.

Symmetry operators for the conformal Laplacian~\cite{km}, Maxwell's
equations~\cite{kmw}, and the Dirac operator~\cite{msw} have been much studied
in general relativity. This is owing to the separation of variables that they
induce. These matters are
discussed further in~\S\ref{discuss}.

This article is the result of questions and suggestions from Edward Witten. In
particular, he suggested that Theorems~\ref{Killing} and~\ref{existence} should
be true and that they lead to an understanding of the symmetry algebra. For
this, and other help, I am extremely grateful. I would also like to thank Erik
van den Ban, Andreas \v{C}ap, Rod Gover, Robin Graham, Bertram Kostant, Toshio
Oshima, Paul Tod, and Joseph Wolf for useful conversations.

\section{Notation and Statement of Results}\label{statement}
Sometimes we shall work on a Riemannian manifold, in which case $\nabla_a$ will
denote the metric connection. Mostly, we shall be concerned with Euclidean
space ${\mathbb R}^n$ and then $\nabla_a=\partial/\partial x^a$,
differentiation in co\"ordinates. In any case, we shall adopt the standard
convention of raising and lowering indices with the metric $g_{ab}$. Thus,
$\nabla^a=g^{ab}\nabla_b$ and $\Delta=\nabla^a\nabla_a$ is the Laplacian. Here
and throughout, we employ the Einstein summation convention: repeated indices
carry an implicit sum. The use of of indices does not refer to any particular
choice of co\"ordinates. Indices are merely markers, serving to identify the
type of tensor under consideration. Formally, this is Penrose's abstract index
notation~\cite{OT}.

We shall be working on Euclidean space ${\mathbb R}^n$
or on a Riemannian manifold of dimension~$n$. We shall always suppose that
$n\geq 3$ (ensuring that the space of conformal Killing vectors is
finite-dimensional).

Kostant~\cite{kos} considers first order linear differential operators
${\mathcal D}$ such that $[{\mathcal D},\Delta]=h\Delta$ for some function~$h$.
We extend these considerations to higher order operators:--
\begin{defn} A symmetry of the Laplacian is a linear differential
\mbox{operator} ${\mathcal D}$ so that $\Delta{\mathcal D}=\delta\Delta$ for
some linear differential operator~$\delta$. \end{defn}
\noindent In particular, such a symmetry preserves harmonic functions. A
rather trivial way in which ${\mathcal D}$ may be a symmetry of the Laplacian
is if it is of the form ${\mathcal P}\Delta$ for some linear differential
operator~${\mathcal P}$. Such an operator kills harmonic functions. In order to
suppress such trivialities, we shall say that two symmetries of the Laplacian
${\mathcal D}_1$ and ${\mathcal D}_2$ are equivalent if and only if
${\mathcal D}_1-{\mathcal D}_2={\mathcal P}\Delta$ for some~${\mathcal P}$. It
is evident that symmetries of the Laplacian are closed under composition and
that composition respects equivalence. Thus, we have an algebra:--
\begin{defn}\label{definitionofAn} The symmetry algebra ${\mathcal A}_n$
comprises symmetries of the Laplacian on ${\mathbb R}^n$, considered up to
equivalence, with algebra operation induced by composition. \end{defn}
\noindent The aim of this article is to study this algebra. We shall also be
able to say something about the corresponding algebra on a Riemannian manifold.
The signature of the metric is irrelevant. All results have obvious
counterparts in the pseudo-Riemannian setting. On Minkowski space, for
example, these counterparts are concerned with symmetries of the wave operator.

Any linear differential operator on a Riemannian manifold may be written in
the form
$${\mathcal D}=
V^{bc\cdots d}\nabla_b\nabla_c\cdots\nabla_d+\mbox{ lower order terms},$$
where $V^{bc\cdots d}$ is symmetric in its indices. This tensor is called the
symbol of~${\mathcal D}$. We shall write $\phi^{(ab\cdots c)}$ for the
symmetric part of $\phi^{ab\cdots c}$.

\begin{defn} A conformal Killing tensor
is a symmetric trace-free tensor field with $s$ indices satisfying
\begin{equation}\label{Killingequation}
\mbox{\rm the trace-free part of }\nabla^{(a}V^{bc\cdots d)}=0\end{equation}
or, equivalently,
\begin{equation}\label{killingtensor}
\nabla^{(a}V^{bc\cdots d)}=g^{(ab}\lambda^{c\cdots d)}\end{equation}
for some tensor field $\lambda^{c\cdots d}$ or, equivalently,
\begin{equation}\label{explicitKilling}
\nabla^{(a}V^{bc\cdots d)}=\textstyle\frac{s}{n+2s-2}
g^{(ab}\nabla_eV^{c\cdots d)e}.\end{equation}
\end{defn}
\noindent When $s=1$, these equations define a conformal
Killing vector.

\begin{thm}\label{Killing} Any symmetry ${\mathcal D}$ of the Laplacian on a
Riemannian manifold is canonically equivalent to one whose
symbol is a conformal Killing tensor. \end{thm}
\begin{proof}Since
$$g^{(bc}\mu^{d\cdots e)}\nabla_b\nabla_c\nabla_d\cdots\nabla_e=
\mu^{d\cdots e}\nabla_d\cdots\nabla_e\Delta+\mbox{ lower order terms},$$
any trace in the symbol of ${\mathcal D}$ may be canonically removed by using
equivalence. Thus, let us suppose that
$${\mathcal D}=V^{bcd\cdots e}\nabla_b\nabla_c\nabla_d\cdots\nabla_e
+\mbox{ lower order terms}$$
is a symmetry of $\Delta$ and that $V^{bcd\cdots e}$ is trace-free symmetric.
Then
$$\begin{array}{rcr}\Delta{\mathcal D}&=&
V^{bcd\cdots e}\nabla_b\nabla_c\nabla_d\cdots\nabla_e\Delta+
2\nabla^{(a}V^{bcd\cdots e)}
  \nabla_a\nabla_b\nabla_c\nabla_d\cdots\nabla_e\quad\\[3pt]
&&+\mbox{ lower order terms}
\end{array}$$
and the only way that the Laplacian can emerge from the sub-leading term is if
(\ref{killingtensor}) holds.
\end{proof}

\begin{thm}\label{existence}
Suppose $V^{b\cdots c}$ is a conformal Killing tensor on ${\mathbb R}^n$ with
$s$ indices. Then there are canonically defined differential operators
${\mathcal D}_V$ and $\delta_V$ each having $V^{b\cdots c}$ as their symbol so
that $\Delta{\mathcal D}_V=\delta_V\Delta$.
\end{thm}
\noindent We shall prove this Theorem in the following section but here are
some examples. When $s=1$,
\begin{equation}\label{examplewhensisone}\begin{array}l
{\mathcal D}_Vf=V^a\nabla_af+\frac{n-2}{2n}(\nabla_aV^a)f\\[7pt]
\delta_Vf=V^a\nabla_af+\frac{n+2}{2n}(\nabla_aV^a)f.\end{array}
\end{equation}
When $s=2$,
\begin{equation}\label{examplewhensistwo}\begin{array}l{\mathcal D}_Vf=
V^{ab}\nabla_a\nabla_bf+\frac{n}{n+2}(\nabla_aV^{ab})\nabla_bf+
\frac{n(n-2)}{4(n+2)(n+1)}(\nabla_a\nabla_bV^{ab})f\\[7pt]
\delta_Vf=V^{ab}\nabla_a\nabla_bf+\frac{n+4}{n+2}(\nabla_aV^{ab})\nabla_bf+
\frac{n+4}{4(n+1)}(\nabla_a\nabla_bV^{ab})f.\end{array}\end{equation}
On~${\mathbb R}^n$, we shall write down in \S\ref{flatresults} all solutions of
the conformal Killing equation~(\ref{killingtensor}). For tensors with $s$
indices, these solutions form a finite-dimensional vector space
${\mathcal K}_{n,s}$ of dimension
\begin{equation}\label{dimension}
\frac{(n+s-3)!(n+s-2)!(n+2s-2)(n+2s-1)(n+2s)}{s!(s+1)!(n-2)!n!}.\end{equation}
Therefore, Theorem~\ref{existence} shows the existence of many symmetries of
the Laplacian on~${\mathbb R}^n$. Together with Theorem~\ref{Killing}, it also
allows us to put any symmetry into a canonical form. Specifically, if
${\mathcal D}$ is a symmetry operator of order~$s$, then we may
apply Theorem~\ref{Killing} to normalise its symbol $V^{b\cdots c}$ to be a
conformal Killing tensor. Now consider ${\mathcal D}-{\mathcal D}_V$ where
${\mathcal D}_V$ is from Theorem~\ref{existence}. By construction, this is a
symmetry of the Laplacian order less than~$s$. Continuing in this fashion we
obtain a canonical form, namely
$${\mathcal D}_{V_s}+{\mathcal D}_{V_{s-1}}+\cdots+{\mathcal D}_{V_2}
+{\mathcal D}_{V_1}+V_0,$$
where $V_t$ is a conformal Killing tensor with $t$ indices (whence $V_1$ is a
conformal Killing vector and $V_0$ is constant). As a vector space, therefore,
Theorems~\ref{Killing} and~\ref{existence} imply a canonical isomorphism
$${\mathcal A}_n=\bigoplus_{s=0}^\infty\mathcal K_{n,s}\,.$$
In the following section, we shall identify ${\mathcal K}_{n,s}$ more
explicitly. This will enable us, in~\S\ref{algebraicstructure}, to prove the
following theorem identifying the algebraic structure on~${\mathcal A}_n$. To
state it, we need some notation. If we identify
${\mathfrak{so}}(n+1,1)=\bigwedge^2{\mathbb R}^{n+2}$, then $V\wedge W$ is an
irreducible component of the symmetric tensor product $V\odot W$, for
$V,W\in{\mathfrak{so}}(n+1,1)$. Let $V\topten W$ denote the trace-free part of
$V\odot W-V\wedge W$.
\begin{thm}\label{structure} The algebra ${\mathcal A}_n$ is isomorphic to the
tensor algebra
$$\mbox{\large$\displaystyle \bigoplus_{s=0}^\infty
\textstyle\bigotimes^s$}{\mathfrak{so}}(n+1,1)$$
modulo the two-sided ideal generated by the elements
\begin{equation}\label{generators}V\otimes W-V\topten W-\frac{1}{2}[V,W]
+\frac{n-2}{4(n+1)}\langle V,W\rangle\end{equation}
for $V,W\in{\mathfrak{so}}(n+1,1)$.
\end{thm}
\noindent Here, $[V,W]$ denotes the Lie bracket of $V$ and $W$ and
$\langle V,W\rangle$ their inner product with respect to the Killing form (as
normalised in~\S\ref{algebraicstructure}).
We can rewrite Theorem~\ref{structure} as saying that ${\mathcal A}_n$ is the
associative algebra generated by ${\mathfrak{so}}(n+1,1)$ but subject to the
relations:--
$$VW-WV=[V,W]\quad\mbox{and}\quad
  VW+WV=2V\topten W-\frac{n-2}{2(n+1)}\langle V,W\rangle.$$
In other words, we have the following description of~${\mathcal A}_n$.
\begin{cor}\label{Aasalgebra}The algebra ${\mathcal A}_n$ is isomorphic to the
enveloping algebra ${\mathfrak U}({\mathfrak{so}}(n+1,1))$ modulo the
two-sided ideal generated by the elements
$$VW+WV-2V\topten W+\frac{n-2}{2(n+1)}\langle V,W\rangle$$
for $V,W\in{\mathfrak{so}}(n+1,1)$.
\end{cor}

In \S\ref{curved} we shall work on a general curved background and prove the
following result.
\begin{thm}\label{betterthanexistence}
Suppose $V^{b\cdots c}$ is a trace-free symmetric tensor field with $s$ indices
on a conformal manifold. Then, for any $w\in{\mathbb R}$, there is a naturally
defined, conformally invariant differential operator, taking densities of
weight $w$ to densities of the same weight~$w$, and having $V^{b\cdots c}$ as
its symbol. If the background metric is flat, $w=1-n/2$, and $V^{b\cdots c}$ is
a conformal Killing tensor, then ${\mathcal D}_V$ agrees with the symmetry
operator given in Theorem~\ref{existence}.
\end{thm}
\noindent When $s=2$, for example,
$$\begin{array}l{\mathcal D}_Vf=
V^{ab}\nabla_a\nabla_bf-\frac{2(w-1)}{n+2}(\nabla_aV^{ab})\nabla_bf\\[5pt]
\phantom{{\mathcal D}_Vf=V^{ab}\nabla_a\nabla_bf}+
\frac{w(w-1)}{(n+2)(n+1)}(\nabla_a\nabla_bV^{ab})f+
\frac{w(n+w)}{(n+1)(n-2)}R_{ab}V^{ab}f,\end{array}$$
where $R_{ab}$ is the Ricci tensor. This extends
(\ref{examplewhensistwo}) to the curved setting.

\section{Results in the Flat Case}\label{flatresults}
The proof of Theorem~\ref{existence} is best approached in the realm of
conformal geometry. As detailed in~\cite[\S{2}]{eg}, ${\mathbb R}^n$ may be
conformally compactified as the sphere $S^n\subset{\mathbb{RP}}_{n+1}$
of null directions of the indefinite quadratic form
\begin{equation}\label{quadraticform}
\widetilde g_{AB}x^Ax^B=2x^0x^\infty+g_{ab}x^ax^b
\quad\mbox{for }x^A=(x^0,x^a,x^\infty)\end{equation}
on~${\mathbb R}^{n+2}$. Then, the conformal symmetries of $S^n$ are induced by
the action of ${\mathrm{SO}}(n+1,1)$ on ${\mathbb R}^{n+2}$ realised as those
linear transformations preserving (\ref{quadraticform}) and of unit
determinant.

We need to incorporate the Laplacian into this picture. To do so,
suppose $F$ is a smooth function defined in a neighbourhood of the
origin in~${\mathbb R}^n$. Then, for any $w\in{\mathbb R}$,
$$f(x^0,x^0x^a,-x^0x^ax_a/2)=(x^0)^wF(x^a)\quad\mbox{for }x^0>0$$
defines a smooth function $f$ on a conical neighbourhood of $(1,0,0)$ in the
null cone ${\mathcal N}$ of the quadratic form~(\ref{quadraticform}). This
is a homogeneous function of degree~$w$, namely
$f(\lambda x^A)=\lambda^wf(x^A)$, for $\lambda>0$.
Conversely, $F$ may be recovered from $f$ by setting $x^0=1$. Hence,
for fixed $w$, the functions $F$ and $f$ are equivalent. In the language of
conformal differential geometry, $w$ is the {\em conformal weight\/} of $F$
when viewed on ${\mathcal N}$ in this way.

Let $\widetilde\Delta$ denote the ambient wave operator
$$\widetilde\Delta=
\widetilde g^{AB}\frac{\partial^2}{\partial x^A\partial x^B}$$
where $\widetilde g^{AB}$ is the inverse of $\widetilde g_{AB}$. Let
$r=\widetilde g_{AB}x^Ax^B$. Then \mbox{${\mathcal N}=\{r=0\}$}. Now consider
$f$, homogeneous of degree $w$ near $(1,0,0)\in{\mathcal N}$. Choose a smooth
ambient extension $\widetilde f$ of $f$ as a homogeneous function defined near
$(1,0,0)\in{\mathbb R}^{n+2}$. Any other such extension will have the form
$\widetilde f+rg$ where $g$ is homogeneous of degree $w-2$. A simple
calculation gives
$$\widetilde\Delta(rg)=r\widetilde\Delta g+2(n+2w-2)g.$$
It follows immediately that, if $w=1-n/2$, then
$\widetilde\Delta\widetilde f|_{\mathcal N}$ depends only on~$f$. This defines
a differential operator on ${\mathbb R}^n$ and, as detailed in~\cite{eg}, one
may easily verify that it is the Laplacian. The main point of this construction
is that it is manifestly invariant under the action of ${\mathrm{SO}}(n+1,1)$.
We say that $\Delta$ is conformally invariant acting on conformal densities of
weight $1-n/2$ on ${\mathbb R}^n$.

This construction is due to Hughston and Hurd~\cite{hh}. Fefferman and
Graham~\cite{fg} extended the argument to general Riemannian manifolds,
producing the conformal Laplacian or Yamabe operator
\begin{equation}\label{yamabe}\Delta-\frac{n-2}{4(n-1)}R,\end{equation}
where $R$ is scalar curvature. Their construction is an early form of the
AdS/CFT correspondence~\cite{ma,wi}. Many other conformally invariant
differential operators were constructed in this manner by Jenne~\cite{j}.
Arbitrary powers of the Laplacian $\Delta^k$ are conformally invariant, in the
flat case, when acting on densities of weight $k-n/2$. This is demonstrated in
\cite[Proposition~4.4]{eg} by an ambient argument.

Conformal Killing tensors have a simple ambient interpretation. This is to be
expected since the equation (\ref{Killingequation}) is conformally invariant.
In fact, the differential operator that is the left hand side of
(\ref{Killingequation}) is the first operator in a conformally invariant
complex of operators known as the Bernstein-Gelfand-Gelfand
complex~\cite{beastwood,bgg,cd,css,l}. This implies that the conformal Killing
tensors on ${\mathbb R}^n$ form an irreducible representation of the conformal
Lie algebra ${\mathfrak{so}}(n+1,1)$, namely
$$\underbrace{\begin{picture}(100,30)
\put(0,5){\line(1,0){100}}
\put(0,15){\line(1,0){100}}
\put(0,25){\line(1,0){100}}
\put(0,5){\line(0,1){20}}
\put(10,5){\line(0,1){20}}
\put(20,5){\line(0,1){20}}
\put(30,5){\line(0,1){20}}
\put(40,5){\line(0,1){20}}
\put(70,5){\line(0,1){20}}
\put(80,5){\line(0,1){20}}
\put(90,5){\line(0,1){20}}
\put(100,5){\line(0,1){20}}
\put(55,10){\makebox(0,0){$\cdots$}}
\put(55,20){\makebox(0,0){$\cdots$}}
\put(105,7){\makebox(0,0)[l]{\scriptsize trace-free part}}
\end{picture}}_{\mbox{$s$ boxes in each row}}$$
as a Young tableau. This is the vector space that we earlier denoted
by~${\mathcal K}_{n,s}$. The formula (\ref{dimension}) for its dimension is
easily obtained from~\cite{ki}. It is convenient to adopt a realisation
of this representation as tensors
$$\textstyle V^{BQCR\cdots DS}\in\bigotimes^{2s}{\mathbb R}^{n+2}$$
that are skew in each pair of indices $BQ$, $CR$, \ldots, $DS$, are totally
trace-free, and so that skewing over any three indices gives zero. (It follows
that $V^{BQCR\cdots DS}$ is symmetric in the paired indices and that
symmetrising over any $s+1$ indices gives zero.) When $s=1$, for example, we
have
$$\textstyle V^{BQ}\in\bigwedge^2{\mathbb R}^{n+2}={\mathfrak{s0}}(n+1,1).$$
This is the well-known identification of conformal Killing vectors as elements
of the conformal Lie algebra. More specifically, following the conventions
of ~\cite{eg}, we have
$$V^B{}_Q=\left\lgroup\begin{array}{ccc}V^0{}_0&V^0{}_q&V^0{}_\infty\\[3pt]
V^b{}_0&V^b{}_q&V^b{}_\infty\\[3pt]
V^\infty{}_0&V^\infty{}_q&V^\infty{}_\infty\end{array}\right\rgroup=
\left\lgroup\begin{array}{ccc}\lambda&r_q&0\\[3pt]
s^b&m^b{}_q&-r^b\\[3pt]
0&-s_q&-\lambda\end{array}\right\rgroup$$
and corresponds to the conformal Killing vector
$$V^b=-s^b-m^b{}_qx^q+\lambda x^b+r_qx^qx^b-(1/2)x_qx^qr^b.$$
More succinctly, if we introduce
$$\Phi^B=\left\lgroup\begin{array}c1\\ x^b\\ -x^bx_b/2\end{array}\right\rgroup
\quad\mbox{and}\quad
\Psi^{bQ}=\left\lgroup\begin{array}c0\\ g^{bq}\\ -x^b\end{array}\right\rgroup,
$$
then, using the ambient metric $\widetilde g_{AB}$ to lower indices,
$$V^{BQ}\mapsto V^b=\Phi_BV^{BQ}\Psi^b{}_Q$$
associates to the ambient skew tensor $V^{BQ}$, the corresponding Killing
vector~$V^b$. This formula immediately generalises:--
$$V^{BQCR\cdots DS}\mapsto V^{bc\cdots d}=\Phi_B\Phi_C\cdots\Phi_D
  V^{BQCR\cdots DS}\Psi^b{}_Q\Psi^c{}_R\cdots\Psi^d{}_S.$$
It is readily verified that if $V^{BQCR\cdots DS}$ satisfies the
symmetries listed above, then $V^{bc\cdots d}$ is trace-free symmetric and
satisfies the conformal Killing equation~(\ref{Killingequation}).
\begin{prop}\label{generalsolution}
This gives the general conformal Killing tensor.\end{prop}
\begin{proof} This is a special case of Lepowsky's generalisation \cite{l} of
the Bernstein-Gelfand-Gelfand resolution. A direct proof may be gleaned
from~\cite{g}. \end{proof}
\renewcommand{\proofname}{Proof of Theorem~\ref{existence}}\begin{proof}
We are now in a position to prove this theorem by ambient methods. Let
$\partial_A$ denote the ambient derivative $\partial/\partial x^A$
on~${\mathbb R}^{n+2}$ and for $V^{BQCR\cdots DS}$ as above, consider the
differential operator $${\mathfrak D}_V=V^{BQCR\cdots DS}
x_Bx_C\cdots x_D\partial_Q\partial_R\cdots\partial_S$$
on ${\mathbb R}^{n+2}$. Evidently, ${\mathfrak D}_V$ preserves homogeneous
functions. Recall that $r=x^Ax_A$. Using $\partial_Ar=2x_A$, it follows that
\begin{equation}\label{ambientD}
{\mathfrak D}_V(rg)=r{\mathfrak D}_Vg\quad\mbox{and}\quad
\widetilde\Delta{\mathfrak D}_V={\mathfrak D}_V\widetilde\Delta.\end{equation}
The first of these implies that ${\mathfrak D}_V$ induces differential
operators on ${\mathbb R}^n$ for densities of any conformal weight: simply
extend the corresponding homogeneous function on ${\mathcal N}$ into
${\mathbb R}^{n+2}$, apply ${\mathfrak D}_V$, and restrict back to~${\mathcal
N}$. In particular, let us denote by ${\mathcal D}_V$ and $\delta_V$ the
differential operators so induced on densities of weight $1-n/2$ and $-1-n/2$,
respectively. Bearing in mind the ambient construction of the Laplacian, it
follows immediately from the second equation of (\ref{ambientD}) that
$\Delta{\mathcal D}_V=\delta_V\Delta$. It remains to calculate the symbols
of ${\mathcal D}_V$ and~$\delta_V$.

To do this first note that, by construction, their order is at most~$s$. For
any such operator ${\mathcal D}$, the symbol at $y\in{\mathbb R}^n$ is given by
$$\left.{\mathcal D}\:
  \frac{(x^b-y^b)(x^c-y^c)\cdots (x^d-y^d)}{s!}\right|_{x=y}.$$
This is easily computed. As a homogeneous function of degree $w$
on~${\mathcal N}$, the function $x^b-y^b$ may be ambiently extended as
$$(x^0,x^a,x^\infty)\mapsto (x^0)^{w-1}x^b-(x^o)^wy^b.$$
Then,
$$\partial^Q((x^0)^{w-1}x^b-(x^o)^wy^b)=
\left\lgroup\begin{array}c0\\ (x^0)^{w-1}g^{bq}\\
(w-1)(x^0)^{w-2}x^b-w(x^0)^{w-1}y^b\end{array}\right\rgroup$$
and when $x^0=1$ and $x=y$, this becomes $\Psi^{bQ}$ at~$y$. Similarly, $x^B$
becomes $\Phi^B$ and, in case $s=1$, we obtain $\Phi_BV^{BQ}\Psi^b{}_Q$. In
other words, the symbol is $V^b$ no matter what is the weight.
The case of general $s$ is similar.
\end{proof}\renewcommand{\proofname}{Proof}
Notice that, not only have we proved Theorem~\ref{existence}, but also we
have a very simple ambient construction of the symmetries~${\mathcal D}_V$.
Explicit formulae for ${\mathcal D}_V$ are another matter. Such formulae can,
of course, be derived from the ambient construction but an easier route will be
provided in ~\S\ref{curved}.

\section{The algebraic structure of ${\mathcal A}_n$}
\label{algebraicstructure}
In view of Theorem~\ref{existence}, Proposition~\ref{generalsolution}, and the
discussion in~\S\ref{statement}, we have identified ${\mathcal A}_n$ as a
vector space:--
\begin{equation}\label{Aasvectorspace}\raisebox{5pt}
{\Large $\displaystyle{\mathcal A}_n\cong\bigoplus_{s=0}^\infty$}\;
\underbrace{\begin{picture}(100,30)
\put(0,5){\line(1,0){100}}
\put(0,15){\line(1,0){100}}
\put(0,25){\line(1,0){100}}
\put(0,5){\line(0,1){20}}
\put(10,5){\line(0,1){20}}
\put(20,5){\line(0,1){20}}
\put(30,5){\line(0,1){20}}
\put(40,5){\line(0,1){20}}
\put(70,5){\line(0,1){20}}
\put(80,5){\line(0,1){20}}
\put(90,5){\line(0,1){20}}
\put(100,5){\line(0,1){20}}
\put(55,10){\makebox(0,0){$\cdots$}}
\put(55,20){\makebox(0,0){$\cdots$}}
\put(105,7){\makebox(0,0)[l]{\scriptsize trace-free part}}
\end{picture}}_{\mbox{$s$}}\end{equation}
but we have yet to identify ${\mathcal A}_n$ as an associative algebra. To do
this, let us first consider the composition ${\mathcal D}_V{\mathcal D}_W$ in
case $V,W\in{\mathfrak{so}}(n+1,1)$. As ambient tenors, $V^{BQ}$ and $V^{CR}$
are skew. {From} the proof of Theorem~\ref{existence}, the operators
${\mathcal D}_V$ and ${\mathcal D}_W$ on ${\mathbb R}^n$ are induced by the
ambient operators
$${\mathfrak D}_V=V^{BQ}x_B\partial_Q\quad\mbox{and}\quad
{\mathfrak D}_W=W^{CR}x_C\partial_R,$$
respectively. Their composition is, therefore, induced by
\begin{equation}\label{compo}{\mathfrak D}_V{\mathfrak D}_W=
V^{BQ}W^{CR}x_Bx_C\partial_Q\partial_R+V^B{}_CW^{CR}x_B\partial_R.
\end{equation}
If we write
$$V^{BQ}W^{CR}=T^{BQCR}+\widetilde g^{BC}U^{QR}+\widetilde g^{QR}U^{BC}
-\widetilde g^{QC}U^{BR}-\widetilde g^{BR}U^{QC}$$
where
\begin{equation}\label{thisisU}U^{BR}=-\frac{1}{n}V^B{}_CW^{CR}
+\frac{1}{2n(n+1)}V^Q{}_CW^C{}_Q\widetilde g^{BR},\end{equation}
then it is easy to verify that $T^{BQCR}$ is totally trace-free.
Now, from~(\ref{compo}), we may rewrite
$$\begin{array}l{\mathfrak D}_V{\mathfrak D}_W=
T^{BQCR}x_Bx_C\partial_Q\partial_R
+rU^{QR}\partial_Q\partial_R+U^{BC}x_Bx_C\widetilde\Delta\\[3pt]
\phantom{{\mathfrak D}_V{\mathfrak D}_W=}
-U^{BR}x_Bx^C\partial_C\partial_R-U^{QC}x_Cx^B\partial_B\partial_Q
+V^B{}_CW^{CR}x_B\partial_R
\end{array}$$
and, bearing in mind (\ref{thisisU}) and that $x^C\partial_C$ is the Euler
operator, if $f$ has homogeneity $w$, then
$$\begin{array}l{\mathfrak D}_V{\mathfrak D}_Wf=
T^{BQCR}x_Bx_C\partial_Q\partial_Rf
+rU^{QR}\partial_Q\partial_Rf+U^{BC}x_Bx_C\widetilde\Delta f\\[5pt]
\phantom{{\mathfrak D}_V{\mathfrak D}_Wf=}\displaystyle
+\frac{w-1+n}{n}V^B{}_CW^{CR}x_B\partial_Rf
+\frac{w-1}{n}V^R{}_CW^{CB}x_B\partial_Rf\\[9pt]
\phantom{{\mathfrak D}_V{\mathfrak D}_Wf=}\displaystyle
-\frac{w(w-1)}{n(n+1)}V^R{}_CW^C{}_Rf.
\end{array}$$
In particular, if $w=1-n/2$, then
$$\begin{array}l{\mathfrak D}_V{\mathfrak D}_Wf=
T^{BQCR}x_Bx_C\partial_Q\partial_Rf
+rU^{QR}\partial_Q\partial_Rf+U^{BC}x_Bx_C\widetilde\Delta f\\[5pt]
\phantom{X}\displaystyle+\frac{1}{2}V^B{}_CW^{CR}x_B\partial_Rf
-\frac{1}{2}V^R{}_CW^{CB}x_B\partial_Rf-\frac{n-2}{4(n+1)}V^R{}_CW^C{}_Rf.
\end{array}$$
Noting that $x_Bx_C\partial_Q\partial_R$ is symmetric in $BC$ and~$QR$, we may
rewrite the first term and obtain, upon restriction to~${\mathcal N}$,
\begin{equation}\label{relations}
{\mathcal D}_V{\mathcal D}_W\equiv{\mathcal D}_{(VW)_2}+{\mathcal D}_{(VW)_1}
+{\mathcal D}_{(VW)_0}\mod\Delta,\end{equation}
as predicted by Theorems~\ref{Killing} and \ref{existence}, where
$$\begin{array}{rcl}
(VW)_2{}^{BQCR}&=&(T^{BQCR}+T^{CRBQ})/3\\
                &&\phantom{XXX}+(T^{BCQR}+T^{QRBC}-T^{QCBR}-T^{BRQC})/6\\[3pt]
(VW)_1{}^{BQ}&=&(V^B{}_CW^{CQ}-V^Q{}_CW^{CB})/2\\[3pt]
(VW)_0&=&-(n-2)V^R{}_CW^C{}_R/(4(n+1)).
\end{array}$$
Each of these expressions has a simple interpretation as follows. The first of
them is the highest weight part of $V\otimes W$. More specifically,
$\bigotimes^2{\mathfrak {so}}(n+1,1)$ decomposes into six irreducibles:--
\begin{equation}\label{decompose}
\raisebox{-25pt}{$\begin{picture}(10,50)(0,-20)
\put(0,5){\line(1,0){10}}
\put(0,15){\line(1,0){10}}
\put(0,25){\line(1,0){10}}
\put(0,5){\line(0,1){20}}
\put(10,5){\line(0,1){20}}
\end{picture}
\raisebox{31pt}{$\;\otimes\;$}
\begin{picture}(10,50)(0,-20)
\put(0,5){\line(1,0){10}}
\put(0,15){\line(1,0){10}}
\put(0,25){\line(1,0){10}}
\put(0,5){\line(0,1){20}}
\put(10,5){\line(0,1){20}}
\end{picture}
\raisebox{31pt}{$\;=\;$}
\begin{picture}(25,50)(0,-20)
\put(0,5){\line(1,0){20}}
\put(0,15){\line(1,0){20}}
\put(0,25){\line(1,0){20}}
\put(0,5){\line(0,1){20}}
\put(10,5){\line(0,1){20}}
\put(20,5){\line(0,1){20}}
\put(22,7){\makebox(0,0)[l]{$\circ$}}
\end{picture}
\raisebox{31pt}{$\;\oplus\;$}
\begin{picture}(27,50)(0,-20)
\put(0,10){\line(1,0){20}}
\put(0,20){\line(1,0){20}}
\put(0,10){\line(0,1){10}}
\put(10,10){\line(0,1){10}}
\put(20,10){\line(0,1){10}}
\put(22,12){\makebox(0,0)[l]{$\circ$}}
\end{picture}
\raisebox{31pt}{$\;\oplus\;{\mathbb R}\;\oplus\;$}
\begin{picture}(20,50)
\put(0,15){\line(1,0){10}}
\put(0,25){\line(1,0){10}}
\put(0,35){\line(1,0){20}}
\put(0,45){\line(1,0){20}}
\put(0,15){\line(0,1){30}}
\put(10,15){\line(0,1){30}}
\put(20,35){\line(0,1){10}}
\put(12,17){\makebox(0,0)[l]{$\circ$}}
\end{picture}
\raisebox{31pt}{$\;\oplus\;$}
\begin{picture}(10,50)(0,-20)
\put(0,5){\line(1,0){10}}
\put(0,15){\line(1,0){10}}
\put(0,25){\line(1,0){10}}
\put(0,5){\line(0,1){20}}
\put(10,5){\line(0,1){20}}
\end{picture}
\raisebox{31pt}{$\;\oplus\;$}
\begin{picture}(10,50)
\put(0,5){\line(1,0){10}}
\put(0,15){\line(1,0){10}}
\put(0,25){\line(1,0){10}}
\put(0,35){\line(1,0){10}}
\put(0,45){\line(1,0){10}}
\put(0,5){\line(0,1){40}}
\put(10,5){\line(0,1){40}}
\end{picture}$}\end{equation}
where $\circ$ denotes the trace-free part. The projection of $V\otimes W$ into
the first of these irreducibles is~$V\topten W$. (More generally, the highest
weight part is known as the Cartan product~\cite[Supplement]{d}.) The
projection
$$\begin{picture}(10,30)
\put(0,5){\line(1,0){10}}
\put(0,15){\line(1,0){10}}
\put(0,25){\line(1,0){10}}
\put(0,5){\line(0,1){20}}
\put(10,5){\line(0,1){20}}
\end{picture}
\raisebox{11pt}{$\;\otimes\;$}
\begin{picture}(10,30)
\put(0,5){\line(1,0){10}}
\put(0,15){\line(1,0){10}}
\put(0,25){\line(1,0){10}}
\put(0,5){\line(0,1){20}}
\put(10,5){\line(0,1){20}}
\end{picture}
\raisebox{11pt}{$\;\ni V^{BQ}W^{CR}\mapsto V^B{}_CW^{CQ}-V^Q{}_CW^{CB}\in\;$}
\begin{picture}(10,30)
\put(0,5){\line(1,0){10}}
\put(0,15){\line(1,0){10}}
\put(0,25){\line(1,0){10}}
\put(0,5){\line(0,1){20}}
\put(10,5){\line(0,1){20}}
\end{picture}$$
is the Lie bracket $V\otimes W\mapsto[V,W]$ and the projection
$$\begin{picture}(10,30)
\put(0,5){\line(1,0){10}}
\put(0,15){\line(1,0){10}}
\put(0,25){\line(1,0){10}}
\put(0,5){\line(0,1){20}}
\put(10,5){\line(0,1){20}}
\end{picture}
\raisebox{11pt}{$\;\otimes\;$}
\begin{picture}(10,30)
\put(0,5){\line(1,0){10}}
\put(0,15){\line(1,0){10}}
\put(0,25){\line(1,0){10}}
\put(0,5){\line(0,1){20}}
\put(10,5){\line(0,1){20}}
\end{picture}
\raisebox{11pt}{$\;\ni V^{BQ}W^{CR}\mapsto V^R{}_CW^C{}_R\in{\mathbb R}$}$$
is the Killing form $V\otimes W\mapsto\langle V,W\rangle$. Thus, we may rewrite
(\ref{relations}) as
\begin{equation}\label{identity}
{\mathcal D}_V{\mathcal D}_W\equiv{\mathcal D}_{V\topten W}
+\frac{1}{2}{\mathcal D}_{[V,W]}
-\frac{n-2}{4(n+1)}{\mathcal D}_{\langle V,W\rangle}\mod\Delta.\end{equation}
In particular, the other irreducibles of (\ref{decompose}) are mapped by
${\mathcal D}$ to zero.
\renewcommand{\proofname}{Proof of Theorem~\ref{structure}}\begin{proof}
Define a map from the tensor algebra to ${\mathcal A}_n$ by
$$V_1\otimes V_2\otimes\cdots\otimes V_s\longmapsto
{\mathcal D}_{V_1}{\mathcal D}_{V_2}\cdots{\mathcal D}_{V_s},$$
extended by linearity. {From} (\ref{identity}) it follows that the
elements (\ref{generators}) are mapped to zero. To complete the proof, it
suffices to consider the corresponding graded algebras. The graded algebra of
${\mathcal A}_n$ is (\ref{Aasvectorspace}) under Cartan product. We must show
that the kernel of the mapping
\begin{equation}\label{proj}
\raisebox{5pt}{$\mbox{\large$\displaystyle \bigoplus_{s=0}^\infty
\textstyle\bigotimes^s$}\;
\begin{picture}(10,30)(0,12)
\put(0,5){\line(1,0){10}}
\put(0,15){\line(1,0){10}}
\put(0,25){\line(1,0){10}}
\put(0,5){\line(0,1){20}}
\put(10,5){\line(0,1){20}}
\end{picture}
\;\longrightarrow\;
\mbox{\large$\displaystyle \bigoplus_{s=0}^\infty$}$}\;
\underbrace{\begin{picture}(100,30)
\put(0,5){\line(1,0){100}}
\put(0,15){\line(1,0){100}}
\put(0,25){\line(1,0){100}}
\put(0,5){\line(0,1){20}}
\put(10,5){\line(0,1){20}}
\put(20,5){\line(0,1){20}}
\put(30,5){\line(0,1){20}}
\put(40,5){\line(0,1){20}}
\put(70,5){\line(0,1){20}}
\put(80,5){\line(0,1){20}}
\put(90,5){\line(0,1){20}}
\put(100,5){\line(0,1){20}}
\put(55,10){\makebox(0,0){$\cdots$}}
\put(55,20){\makebox(0,0){$\cdots$}}
\put(102,7){\makebox(0,0)[l]{$\circ$}}
\end{picture}}_{\mbox{$s$}}\end{equation}
is the two-sided ideal generated by $V\otimes W-V\topten W$ for
$V,W\in\:
\begin{picture}(10,10)(0,8)
\put(0,5){\line(1,0){10}}
\put(0,15){\line(1,0){10}}
\put(0,25){\line(1,0){10}}
\put(0,5){\line(0,1){20}}
\put(10,5){\line(0,1){20}}
\end{picture}\;$.
Equivalently, let us group the decomposition (\ref{decompose}) as
$$\begin{picture}(10,30)
\put(0,5){\line(1,0){10}}
\put(0,15){\line(1,0){10}}
\put(0,25){\line(1,0){10}}
\put(0,5){\line(0,1){20}}
\put(10,5){\line(0,1){20}}
\end{picture}
\raisebox{12pt}{$\;\otimes\;$}
\begin{picture}(10,30)
\put(0,5){\line(1,0){10}}
\put(0,15){\line(1,0){10}}
\put(0,25){\line(1,0){10}}
\put(0,5){\line(0,1){20}}
\put(10,5){\line(0,1){20}}
\end{picture}
\raisebox{12pt}{$\;=\;$}
\begin{picture}(25,30)
\put(0,5){\line(1,0){20}}
\put(0,15){\line(1,0){20}}
\put(0,25){\line(1,0){20}}
\put(0,5){\line(0,1){20}}
\put(10,5){\line(0,1){20}}
\put(20,5){\line(0,1){20}}
\put(22,7){\makebox(0,0)[l]{$\circ$}}
\end{picture}
\raisebox{12pt}{$\;\oplus\;{\mathcal I}_2\,.$}$$
Then ${\mathcal I}_2$ is claimed to generate the kernel of~(\ref{proj}). In
degree $s=2$, this is true by definition.
In degree $s=3$, we must show that
\begin{equation}\label{want}\raisebox{-10pt}{\begin{picture}(10,30)
\put(0,5){\line(1,0){10}}
\put(0,15){\line(1,0){10}}
\put(0,25){\line(1,0){10}}
\put(0,5){\line(0,1){20}}
\put(10,5){\line(0,1){20}}
\end{picture}
\raisebox{12pt}{$\;\otimes\;$}
\begin{picture}(10,30)
\put(0,5){\line(1,0){10}}
\put(0,15){\line(1,0){10}}
\put(0,25){\line(1,0){10}}
\put(0,5){\line(0,1){20}}
\put(10,5){\line(0,1){20}}
\end{picture}
\raisebox{12pt}{$\;\otimes\;$}
\begin{picture}(10,30)
\put(0,5){\line(1,0){10}}
\put(0,15){\line(1,0){10}}
\put(0,25){\line(1,0){10}}
\put(0,5){\line(0,1){20}}
\put(10,5){\line(0,1){20}}
\end{picture}
\raisebox{12pt}{$\;=\;$}
\begin{picture}(35,30)
\put(0,5){\line(1,0){30}}
\put(0,15){\line(1,0){30}}
\put(0,25){\line(1,0){30}}
\put(0,5){\line(0,1){20}}
\put(10,5){\line(0,1){20}}
\put(20,5){\line(0,1){20}}
\put(30,5){\line(0,1){20}}
\put(32,7){\makebox(0,0)[l]{$\circ$}}
\end{picture}
\raisebox{12pt}{$\;\oplus\:\Big(\Big({\mathcal I}_2\:\otimes\;$}
\begin{picture}(10,30)
\put(0,5){\line(1,0){10}}
\put(0,15){\line(1,0){10}}
\put(0,25){\line(1,0){10}}
\put(0,5){\line(0,1){20}}
\put(10,5){\line(0,1){20}}
\end{picture}
\raisebox{12pt}{$\;\Big)\:+\:\Big(\;$}
\begin{picture}(10,30)
\put(0,5){\line(1,0){10}}
\put(0,15){\line(1,0){10}}
\put(0,25){\line(1,0){10}}
\put(0,5){\line(0,1){20}}
\put(10,5){\line(0,1){20}}
\end{picture}
\raisebox{12pt}{$\;\otimes\:{\mathcal I}_2\Big)\Big)\,.$}}\end{equation}
To do this, we first check, by inspection, that
\begin{equation}\label{intersect}\raisebox{-10pt}{\begin{picture}(35,30)
\put(0,5){\line(1,0){30}}
\put(0,15){\line(1,0){30}}
\put(0,25){\line(1,0){30}}
\put(0,5){\line(0,1){20}}
\put(10,5){\line(0,1){20}}
\put(20,5){\line(0,1){20}}
\put(30,5){\line(0,1){20}}
\put(32,7){\makebox(0,0)[l]{$\circ$}}
\end{picture}
\raisebox{12pt}{$\;\;=\:\Big(\;$}
\begin{picture}(25,30)
\put(0,5){\line(1,0){20}}
\put(0,15){\line(1,0){20}}
\put(0,25){\line(1,0){20}}
\put(0,5){\line(0,1){20}}
\put(10,5){\line(0,1){20}}
\put(20,5){\line(0,1){20}}
\put(22,7){\makebox(0,0)[l]{$\circ$}}
\end{picture}
\raisebox{12pt}{$\;\otimes\;$}
\begin{picture}(10,30)
\put(0,5){\line(1,0){10}}
\put(0,15){\line(1,0){10}}
\put(0,25){\line(1,0){10}}
\put(0,5){\line(0,1){20}}
\put(10,5){\line(0,1){20}}
\end{picture}
\raisebox{12pt}{$\;\Big)\:\cap\:\Big(\;$}
\begin{picture}(10,30)
\put(0,5){\line(1,0){10}}
\put(0,15){\line(1,0){10}}
\put(0,25){\line(1,0){10}}
\put(0,5){\line(0,1){20}}
\put(10,5){\line(0,1){20}}
\end{picture}
\raisebox{12pt}{$\;\otimes\;$}
\begin{picture}(25,30)
\put(0,5){\line(1,0){20}}
\put(0,15){\line(1,0){20}}
\put(0,25){\line(1,0){20}}
\put(0,5){\line(0,1){20}}
\put(10,5){\line(0,1){20}}
\put(20,5){\line(0,1){20}}
\put(22,7){\makebox(0,0)[l]{$\circ$}}
\end{picture}
\raisebox{12pt}{$\;\Big)\,.$}}\end{equation}
Specifically, the right hand side consists of tensors $T^{BQCRDS}$ that are
skew in the pairs $BQ$, $CR$, and $DS$, that are trace-free in the indices
$BQCR$ and $CRDS$, and so that skewing over any three indices from $BQCR$ or
$CRDS$ gives zero. In particular, $T^{BQCRDS}$ is symmetric in the paired
indices and, therefore, totally trace-free. To characterise an element of the
left hand side, it remains to show that skewing over $BCD$ gives zero:--
$$\begin{array}l
T^{BQCRDS}+T^{CQDRBS}+T^{DQBRCS}\\
\;-T^{CQBRDS}-T^{DQCRBS}-T^{BQDRCS}\\[3pt]
\;\quad=T^{BCQRDS}+T^{CDQRBS}+T^{DBQRCS}\\[3pt]
\;\qquad=T^{QRBCDS}+T^{QRCDBS}+T^{QRDBCS}=0.\end{array}$$
To continue with (\ref{want}), we may as well establish the corresponding
statement for representations of ${\mathrm{SO}}(n+2)$ since explicit formulae
for decomposing representations of the orthogonal group are independent of the
signature. (Equivalently, we may complexify and employ Weyl's unitary trick.)
Then, the left hand side of (\ref{want}) admits an invariant inner product
with respect to which the projections onto
$$\begin{picture}(25,30)
\put(0,5){\line(1,0){20}}
\put(0,15){\line(1,0){20}}
\put(0,25){\line(1,0){20}}
\put(0,5){\line(0,1){20}}
\put(10,5){\line(0,1){20}}
\put(20,5){\line(0,1){20}}
\put(22,7){\makebox(0,0)[l]{$\circ$}}
\end{picture}
\raisebox{12pt}{$\;\otimes\;$}
\begin{picture}(10,30)
\put(0,5){\line(1,0){10}}
\put(0,15){\line(1,0){10}}
\put(0,25){\line(1,0){10}}
\put(0,5){\line(0,1){20}}
\put(10,5){\line(0,1){20}}
\end{picture}\raisebox{12pt}{\qquad and\qquad}
\begin{picture}(10,30)
\put(0,5){\line(1,0){10}}
\put(0,15){\line(1,0){10}}
\put(0,25){\line(1,0){10}}
\put(0,5){\line(0,1){20}}
\put(10,5){\line(0,1){20}}
\end{picture}
\raisebox{12pt}{$\;\otimes\;$}
\begin{picture}(25,30)
\put(0,5){\line(1,0){20}}
\put(0,15){\line(1,0){20}}
\put(0,25){\line(1,0){20}}
\put(0,5){\line(0,1){20}}
\put(10,5){\line(0,1){20}}
\put(20,5){\line(0,1){20}}
\put(22,7){\makebox(0,0)[l]{$\circ$}}
\end{picture}$$
are orthogonal; let us denote them by $P$ and $Q$, respectively. Then,
(\ref{intersect}) says that
$$\begin{picture}(35,30)
\put(0,5){\line(1,0){30}}
\put(0,15){\line(1,0){30}}
\put(0,25){\line(1,0){30}}
\put(0,5){\line(0,1){20}}
\put(10,5){\line(0,1){20}}
\put(20,5){\line(0,1){20}}
\put(30,5){\line(0,1){20}}
\put(32,7){\makebox(0,0)[l]{$\circ$}}
\end{picture}
\raisebox{12pt}{$\;\;=\im P\:\cap\:\im Q$}$$
and we are required to show that
$$\begin{picture}(10,30)
\put(0,5){\line(1,0){10}}
\put(0,15){\line(1,0){10}}
\put(0,25){\line(1,0){10}}
\put(0,5){\line(0,1){20}}
\put(10,5){\line(0,1){20}}
\end{picture}
\raisebox{12pt}{$\;\otimes\;$}
\begin{picture}(10,30)
\put(0,5){\line(1,0){10}}
\put(0,15){\line(1,0){10}}
\put(0,25){\line(1,0){10}}
\put(0,5){\line(0,1){20}}
\put(10,5){\line(0,1){20}}
\end{picture}
\raisebox{12pt}{$\;\otimes\;$}
\begin{picture}(10,30)
\put(0,5){\line(1,0){10}}
\put(0,15){\line(1,0){10}}
\put(0,25){\line(1,0){10}}
\put(0,5){\line(0,1){20}}
\put(10,5){\line(0,1){20}}
\end{picture}
\raisebox{12pt}{$\;=
\big(\im P \:\cap\:\im Q\big)
\oplus\big(\ker P + \ker Q\big).$}$$
This is a fact concerning orthogonal projections. The composition $QP$
preserves $(\im P\:\cap\:\im Q)^\perp$ and is norm-decreasing there.
Hence, ${\mathrm{Id}}-QP$ is invertible on this subspace. Therefore, we may
write
$$\begin{array}{rcl}T&=&({\mathrm{Id}}-QP)({\mathrm{Id}}-QP)^{-1}T\\[3pt]
&=&\big(({\mathrm{Id}}-P)+({\mathrm{Id}}-Q)P\big)({\mathrm{Id}}-QP)^{-1}T
\end{array}$$
for $T\in(\im P\:\cap\:\im Q)^\perp$, an expression evidently in
$\ker P + \ker Q$. This completes the proof of (\ref{want}) and hence
that~${\mathcal I}_3$, the degree $s=3$ component of the kernel
of~(\ref{proj}), is generated by~${\mathcal I}_2$. Higher components are
similarly dealt with by induction.
\end{proof}\renewcommand{\proofname}{Proof}

\section{Explicit formulae and the curved case}\label{curved}
The ambient construction of ${\mathcal D}_V$ given in the proof of
Theorem~\ref{existence} may be converted into explicit formulae on
${\mathbb R}^n$ using the co\"ordinates (4.4) of~\cite{eg}. It is more
convenient, however, to derive these formulae from their conformal invariance
since, at the same time, we shall obtain conformally invariant differential
operators valid on any Riemannian manifold. We implicitly follow the `Lie
algebra cohomology' method of~\cite[\S{5}]{be}. An alternative approach would
be to follow Gover~\cite{g1,g2}, generalising the ambient operator
$x_B\partial_Q-x_Q\partial_B$ to the curved setting (see also~\cite{cg1,cg2}).
One could also work directly with the ambient metric construction of Fefferman
and Graham~\cite{fg}.

\renewcommand{\proofname}{Proof of Theorem~\ref{betterthanexistence}}
\begin{proof}
We shall follow the conventions of~\cite{beg} concerning
conformal geometry.
On a conformal manifold, we must distinguish \mbox{between}
covariant and contravariant tensors but we can always lower indices if we keep
track of the conformal weight: if $V^a$ has weight~$v$, then $V_a=g_{ab}V^b$
has weight~$v+2$. Unless otherwise stated, we shall suppose that all tensors in
the following discussion are covariant. If $\sigma$ and $\tau$ are symmetric
trace-free tensors, we shall write $\sigma\circ\tau$ to mean the symmetric
trace-free part of $\sigma\otimes\tau$. For example,
$$\textstyle(\sigma\circ\tau)_{abc}=
\sigma_{(a}\tau_{bc)}-\frac{2}{n+2}g_{(ab}\sigma^d\tau_{c)d}.$$
Suppose $\tau^k$ is a symmetric trace-free tensor having $k$ indices and of
conformal weight $w$. Write $\nabla\circ\tau^k$ for the symmetric trace free
part of~$\nabla\tau^k$. Then, under conformal change of metric
$g_{ab}\mapsto\hat g_{ab}=\Omega^2g_{ab}$,
\begin{equation}\label{change}
\hat\nabla\circ\tau^k=\nabla\circ\tau^k+(w-2k)\Upsilon\circ\tau^k,
\end{equation}
where $\Upsilon=d\log\Omega$.
Writing $\Phi_{ab}$ for the trace-free part of~$\frac{1}{n-2}R_{ab}$,
\begin{equation}\label{changeinPhi}
\hat\Phi=\Phi-\nabla\circ\Upsilon+\Upsilon\circ\Upsilon.\end{equation}
Suppose $f$ is a density of conformal weight $w\not\in\{0,1,\ldots,s-1\}$
and consider the symmetric trace-free tensors $\tau^k$ with $k$ indices and of
weight $w$ defined inductively by
\begin{equation}\label{defoftau}
\begin{array}l\tau^0=s!w(w-1)\cdots(w-s+1)f,\\[3pt]
\tau^1=s!(w-1)\cdots(w-s+1)\nabla f,\quad\mbox{and}\\[3pt]
\tau^k
=\frac{1}{k(w-k+1)}\left(\nabla\circ\tau^{k-1}+\Phi\circ\tau^{k-2}\right)
\quad\mbox{for }2\leq k\leq s.\end{array}\end{equation}
For example, if $s=3$, then
\begin{equation}\label{tauincasethree}
\begin{array}{rcl}\tau^0&=&6w(w-1)(w-2)f\\[3pt]
\tau^1&=&6(w-1)(w-2)\nabla f\\[3pt]
\tau^2&=&3(w-2)(\nabla\circ\nabla f+w\Phi\circ f)\\[3pt]
\tau^3&=&
\nabla\circ\nabla\circ\nabla f+(3w-2)\Phi\circ\nabla f+w(\nabla\circ\Phi)f.
\end{array}\end{equation}
These formulae also make sense for $w\in\{0,1,2\}$. More generally, $\tau^k$
are well-defined for all~$w$: the apparently rational coefficients introduced
in their inductive definition are, in fact, polynomial. {From} (\ref{change}),
(\ref{changeinPhi}), and~(\ref{defoftau}), it now follows easily that
\begin{equation}\label{changeintau}
\textstyle \hat\tau^k=\tau^k+\Upsilon\circ\tau^{k-1}
+\frac{1}{2}\Upsilon\circ\Upsilon\circ\tau^{k-2}+\cdots
+\frac{1}{k!}\underbrace{\Upsilon\circ\cdots\circ\Upsilon}_k\circ\tau^0.
\end{equation}

If $\rho$ and $\sigma$ are symmetric trace-free tensors, write
$\rho\intprod\sigma$ to mean the symmetric trace-free tensor obtained by
natural contraction. For example,
$$(\rho\intprod\sigma)_{abc}=\rho^{de}\sigma_{abcde}.$$
Recall that $V^{bc\cdots d}$ is a trace-free symmetric tensor field with $s$
indices. Denote $V_{bc\cdots d}$ by~$\sigma_s$. It has conformal weight~$2s$.
Define
$$\textstyle\sigma_{s-1}=-\frac{1}{n+2s-2}\nabla^b V_{bc\cdots d}
=-\frac{1}{n+2s-2}\nabla\intprod\sigma_s$$
and continue with
\begin{equation}\label{defofsigma}\textstyle\sigma_{s-k}=
-\frac{1}{k(n+2s-k-1)}(\nabla\intprod\sigma_{s-k+1}
-\Phi\intprod\sigma_{s-k+2})\quad\mbox{for }2\leq k\leq s.\end{equation}
Under rescaling of the metric $g_{ab}\mapsto\hat g_{ab}=\Omega^2g_{ab}$,
$$\hat\nabla\intprod\sigma_{s-k}=
\nabla\intprod\sigma_{s-k}+(n+2s-2k-2)\Upsilon\intprod\sigma_{s-k}$$
and it follows that
\begin{equation}\label{changeinsigma}\begin{array}r
\textstyle \hat\sigma_{s-k}=\sigma_{s-k}-\Upsilon\intprod\sigma_{s-k+1}
+\frac{1}{2}(\Upsilon\circ\Upsilon)\intprod\sigma_{s-k+2}+\cdots
\qquad\qquad\\[5pt]
\cdots+
(-1)^k\frac{1}{k!}
\underbrace{(\Upsilon\circ\cdots\circ\Upsilon)}_k\intprod\sigma_0.
\end{array}\end{equation}
{From} (\ref{changeintau}) and (\ref{changeinsigma}) we conclude that
\begin{equation}\label{answer}\sigma_s\intprod\tau^s+
\sigma_{s-1}\intprod\tau^{s-1}\cdots+\sigma_1\intprod\tau^1+\sigma_0\tau^0
\end{equation}
is conformally invariant. By construction, the leading term in $\tau^s$ is
the $s^{\mathrm{th}}$ trace-free symmetric covariant derivative of~$f$.
Therefore, the expression (\ref{answer}) has the form
$$V^{bc\cdots d}\nabla_b\nabla_c\cdots\nabla_df+\mbox{lower order terms, linear
in $V$ and $f$}.$$
This is our definition of ${\mathcal D}_Vf$. It is a conformally invariant
bilinear differential pairing of $V$ and $f$ and is natural in the sense
of~\cite{kms}. It is easily verified that the formulae (\ref{defoftau})
are forced by linearity in~$f$, naturality, and the simple conformal
transformation~(\ref{changeintau}). Similarly, for the collection
$\sigma_s,\ldots,\sigma_0$, linear in~$V$. We conclude that there is no choice
in ${\mathcal D}_Vf$ and, in the flat case, it must agree with the ambient
construction in the proof of Theorem~\ref{existence}.
\end{proof}\renewcommand{\proofname}{Proof}

A side-effect of this proof is the construction of certain
conformally invariant operators. If we set $w=s-1$ in the formulae for
$\tau^k$, only $\tau^s$ is non-zero and (\ref{changeintau}) implies that
$\tau^s$ is conformally invariant. Thus, when $f$ has weight
$w\in{\mathbb Z}_{\geq 0}$, we obtain a conformally invariant operator of the
form
$$f\mapsto\underbrace{\nabla\circ\nabla\circ\cdots\circ\nabla}_{w+1}f
  +\mbox{lower order terms}.$$
When $s=3$, for example, (\ref{tauincasethree}) yields
$$\nabla\circ\nabla\circ\nabla f+4\Phi\circ\nabla f+2(\nabla\circ\Phi)f.$$
acting on $f$ of weight~$2$. The operators in~\cite{dnw} may be constructed by
similar means.

Invariant bilinear differential pairings also appear as the cup product of
Calderbank and Diemer~\cite{cd}. The pairing $(V,f)\mapsto{\mathcal D}_Vf$
of Theorem~\ref{betterthanexistence} is evidently in the same vein but not, in
fact, a special case.

The construction in the proof of Theorem~\ref{betterthanexistence}
gives rise to a formula for ${\mathcal D}_Vf$ in the flat case, namely
$$\sum_{k=0}^s(-1)^{s-k}
\textstyle\binom{s}{k}\frac{(w-s+1)\cdots(w-k)}{(n+2s-2)\cdots(n+s+k-1)}
(\underbrace{\nabla\intprod\cdots\intprod\nabla}_{s-k}\intprod V)\intprod
 \underbrace{\nabla\cdots\nabla}_kf$$
Notice
that, in this formula, there is no need explicitly to remove the trace from
$\nabla\cdots\nabla f$ since it is contracted with
$\nabla\intprod\cdots\intprod\nabla\intprod V$, which is trace-free. The
formulae (\ref{examplewhensisone}) and (\ref{examplewhensistwo}) are special
cases.
\renewcommand{\proofname}{Alternative proof of Theorem~\ref{existence}}
\begin{proof}
The comment just made implies that we can write
${\mathcal D}_Vf=\sum_{k=0}^s\sigma_k\intprod\bar\tau^k$ where, following
(\ref{defoftau}) and (\ref{defofsigma}),
\begin{equation}\label{recurrence}
\begin{array}{rcl}(k+1)(w-k)\bar\tau^{k+1}&=&\nabla\bar\tau^k\\[3pt]
(s-k)(n+s+k-1)\sigma_k&=&-\nabla\intprod\sigma_{k+1}.
\end{array}\end{equation}
We need some differential consequences of $V$ being a conformal Killing
tensor on~${\mathbb R}^n$, specifically that
$$\begin{array}l(s-k+1)(n+s+k-2)
\underbrace{\nabla\intprod\cdots\intprod\nabla}_{s-k}\intprod\Delta V\\[5pt]
=k\Big[(k-1)g\odot
\underbrace{\nabla\intprod\cdots\intprod\nabla}_{s-k+2}\intprod V
-(n+2k-4)\nabla\odot
\underbrace{\nabla\intprod\cdots\intprod\nabla}_{s-k+1}\intprod V
\Big]\end{array}$$
where $\odot$ denotes symmetric tensor product. The case $k=s+1$ is
(\ref{explicitKilling}) and the general case is obtained by taking $s-k+1$
divergences thereof. It now follows from (\ref{recurrence}) that
$$\Delta\sigma_k=k(n+2k-4)\nabla\odot\sigma_{k-1}
+k(k-1)(s-k+2)(n+s+k-3)g\odot\sigma_{k-2}.$$
It is convenient to define $\sigma_{s+1}=0$ so that this equation also holds
when $k=s+1$. At the same time let us define $\bar\tau^{s+1}$
by~(\ref{recurrence}). Bearing in mind that $\nabla\bar\tau^k$ is
automatically symmetric,
$$\Delta{\mathcal D}_Vf=\sum_{k=0}^s\Delta(\sigma_k)\intprod\bar\tau^k
+2(\nabla\odot\sigma_k)\intprod(\nabla\bar\tau^k)
+\sigma_k\intprod\Delta\bar\tau^k$$
and
substituting from (\ref{recurrence}) for~$\nabla\bar\tau^k$, we obtain
$$\begin{array}l\displaystyle\sum_{k=0}^s\Delta(\sigma_k)\intprod\bar\tau^k
+2(k+1)(w-k)(\nabla\odot\sigma_k)\intprod(\bar\tau^{k+1})
+\sigma_k\intprod\Delta\bar\tau^k\\
\displaystyle\qquad=\sum_{k=0}^{s+1}\Delta(\sigma_k)\intprod\bar\tau^k
+2k(w-k+1)(\nabla\odot\sigma_{k-1})\intprod(\bar\tau^k)
+\sigma_k\intprod\Delta\bar\tau^k
\end{array}$$
in which the cross terms cancel provided
$$k(n+2k-4)+2k(w-k+1)=0,\quad\forall k.$$
This is true when $w=1-n/2$, as it should be in the conformally invariant
formulation of Theorem~\ref{existence}. With this value of $w$, we obtain
from~(\ref{recurrence})
$$k(k-1)(n+2k-4)(n+2k-6)g\intprod \bar\tau^k
=4g\intprod\nabla\nabla\bar\tau^{k-2}
=4\Delta\bar\tau^{k-2}$$
and so
\begin{equation}\label{atlasttheanswer}\makebox[0pt]{$
\begin{array}l\displaystyle\Delta{\mathcal D}_Vf=\sum_{k=2}^{s+1}
4\mbox{\small$\displaystyle\frac{(s-k+2)(n+s+k-3)}{(n+2k-4)(n+2k-6)}$}
\sigma_{k-2}\intprod\Delta\bar\tau^{k-2}
+\sum_{k=0}^s\sigma_k\intprod\Delta\bar\tau^k\\
\displaystyle\phantom{\Delta{\mathcal D}_Vf}\qquad=\sum_{k=0}^s
\left(1+
4\mbox{\small$\displaystyle\frac{(s-k)(n+s+k-1)}{(n+2k)(n+2k-2)}$}\right)
\sigma_k\intprod\Delta\bar\tau^k\\
\displaystyle\phantom{\Delta{\mathcal D}_Vf}\qquad\qquad=\sum_{k=0}^s
\mbox{\small$\displaystyle\frac{(n+2s)(n+2s-2)}{(n+2k)(n+2k-2)}$}
\sigma_k\intprod\Delta\bar\tau^k.
\end{array}$}\end{equation}
If $\Delta f=0$, then $\Delta\bar\tau^k=0$ for all $k$ and so
$\Delta{\mathcal D}_Vf=0$, as required. More precisely, the final expression
of (\ref{atlasttheanswer}) is ${\mathcal D}_V$ applied to $\Delta f$, having
conformal weight $-1-n/2$.
\end{proof}\renewcommand{\proofname}{Proof}
This alternative proof, though direct, is a brute force calculation. The
ambient proof given in \S\ref{flatresults} is more conceptual. This is typical
of the AdS/CFT correspondence with effects more clearly visible `in the bulk'.

Explicit formulae allow us, in principle, to calculate the composition
${\mathcal D}_V{\mathcal D}_W$ for conformal Killing tensors $V$ and~$W$, then
to throw the result into canonical form in accordance with
Theorems~\ref{Killing} and~\ref{existence}. In practise, this is difficult but
if $V$ and $W$ are Killing vectors, we find
\begin{equation}\label{compose}
{\mathcal D}_V{\mathcal D}_W={\mathcal D}_{V\topten W}
+\frac{1}{2}{\mathcal D}_{[V,W]}
-\frac{n-2}{4(n+1)}{\mathcal D}_{\langle V,W\rangle}+\frac{1}{n}V^aW_a\Delta,
\end{equation}
where
\begin{equation}\label{pairings}\begin{array}{rcl}
(V\topten W)^{ab}&=&V^{(a}W^{b)}-\frac{1}{n}g^{ab}V^cW_c\\[5pt]
[V,W]^a&=&V^b\nabla_bW^a-W^b\nabla_bV^a\\[5pt]
\langle V,W\rangle&=&(\nabla_bV^a)(\nabla_aW^b)
 -\frac{n-2}{n^2}(\nabla_aV^a)(\nabla_bW^b)\\[3pt]
&&\hspace{50pt}\mbox{ }
 -\frac{2}{n}V^a\nabla_a\nabla_bW^b-\frac{2}{n}W^a\nabla_a\nabla_bV^b.
\end{array}\end{equation}
It is a differential consequence of the conformal Killing equation that
$V\topten W$ is a conformal Killing tensor, $[V,W]$ is a conformal Killing
vector (the usual Lie bracket of vector fields), and $\langle V,W\rangle$ is
constant. Of course, (\ref{compose}) is the form taken by (\ref{identity})
when written on~${\mathbb R}^n$.

\section{Concluding remarks}\label{discuss}
Several questions remain unanswered, the most obvious of which are concerned
with what happens in the curved setting. Though Theorem~\ref{Killing} is stated
for the Laplacian, its proof is equally valid for the conformal
Laplacian~(\ref{yamabe}). The operators of Theorem~\ref{betterthanexistence}
are conformally invariant and natural in the sense of~\cite{kms}. But it is
difficult to say whether they are symmetry operators of the conformal
Laplacian. The conformal Killing equation (\ref{Killingequation}) is
overdetermined and generically has no solutions. Separation of variables for
the geodesic equation was discovered in the Kerr solution by Carter~\cite{car}
and an explanation of this phenomenon in terms of (conformal) Killing tensors
was provided by Walker and Penrose~\cite{wp} (see also~\cite{wo}). In
particular, there are space-times with conformal Killing tensors not arising
from conformal Killing vectors. These can lead to extra symmetries for the
(conformal) Laplacian~\cite{km}. Nevertheless, the relationship to
Theorem~\ref{betterthanexistence}, if any, is unclear.

The algebraic definition of the product on the flat symmetry algebra
${\mathcal A}_n$ is perhaps not as explicit as it might be. According to
(\ref{Aasvectorspace}), as a vector space
$${\mathcal A}_n\cong\raisebox{2pt}{$\displaystyle\bigoplus_{k=0}^\infty$}
\:\mbox{\Large$\topten^k$}({\mathfrak{so}}(n+1,1))$$
where \mbox{\large$\topten$} denotes the Cartan product~\cite{d}. It is
difficult to realise the algebra structure directly from this point of view.
Rather, it is induced by realising ${\mathcal A}_n$ as a quotient of the
universal enveloping algebra of the conformal algebra as in
Corollary~\ref{Aasalgebra}. But, similar comments apply to the universal
enveloping algebra itself. The filtration of ${\mathcal A}_n$ by degree is
induced by the usual filtration on ${\mathfrak U}({\mathfrak{so}}(n+1,1))$ and,
for any Lie algebra~${\mathfrak g}$, there is a canonical vector space
isomorphism with the corresponding graded algebra
${\mathfrak U}({\mathfrak g})\cong\bigodot({\mathfrak g})$, namely the
symmetric tensor algebra of~${\mathfrak g}$. It is possible to transfer the
algebra structure to~$\bigodot({\mathfrak g})$. The result is Kontsevich's
$\star$-product~\cite{ads,kon}.

The algebra structure on ${\mathcal A}_n$ may be expressed in terms of bilinear
differential pairings between conformal Killing tensors. The pairings
(\ref{pairings}) are simple examples. Formulae for the general case are
obscure. The leading term induces an algebraic pairing,
namely the trace-free symmetric product. Also, the first order pairing is
relatively simple: if $V$ has $s$ indices and $W$ has $t$ indices, then
$$V\otimes W\mapsto\mbox{\rm trace-free part of }
\left[sV^{a(b\cdots c}\nabla_aW^{d\cdots f)}
-tW^{a(b\cdots d}\nabla_aV^{e\cdots f)}\right].$$
It is due to Nijenhuis~\cite{n} (with conformal invariance observed by
Woodhouse~\cite{wo}). It is difficult to say whether there are natural higher
order curved pairings.

Finally, we remark that it is straightforward to extend the results of this
article to the two-dimensional case provided the manifold is endowed with a
M\"obius structure in the sense of Calderbank~\cite{cal}.

\end{document}